\documentclass{article}
\usepackage{amsmath,amsthm,amsfonts,graphicx,epsfig,mathrsfs}

\newcommand{\ket}[1]{\left\vert #1\right\rangle}

\begin{document}
\title{A comment on {\it ``A semiconductor source of entangled photons''}}
\author{N. H. Lindner, J. Avron, N. Akopian,  and  D. Gershoni
\\
Department of Physics\\ Technion, 32000 Haifa, Israel}

\maketitle

\begin{abstract} When basic tools of quantum information are applied to the
quantum tomography data presented in \cite{nature}, none of their
devices appears to be a source of entangled photons.
\end{abstract}
In a paper titled {\em ``A semiconductor quantum source of
triggered entangled photon pairs"}  Stevenson {\it et.\ al.}
\cite{nature} claim to find evidence for the emission of
polarization entangled photons from certain quantum dots.

A density matrix completely specifies all properties of the
quantum state \cite{peres_book}.  Using quantum tomography
\cite{kwiat}, the authors of \cite{nature} construct the density
matrices representing the polarization state of the photons
emitted by each of their devices. We show that subjecting
their quantum states to the basic definition of entanglement leads
to the inevitable conclusion that none of their devices produced
entangled light.

In all the dots investigated in \cite{nature}, quantum tomography
yielded real density matrices of the form
\begin{equation}
  \label{eq:density_matrix}
  \rho=
  \left(
  \begin{array}{cccc}
  \alpha & 0 & 0 & \gamma \\
 0 & \beta & 0 & 0 \\
  0 & 0 & \beta'&0 \\
\gamma & 0 & 0 & \alpha' \\
  \end{array}
  \right)\,,
\end{equation}
with $\beta\approx\beta'$ and the zeros stands for terms that are
comparable to the measurements' noise. (The matrix is written in
the basis $\{\ket{HH},\ket{HV},\ket{VH}$, $\ket{VV}\}$.)

A straightforward test for entanglement is the Peres separability
criterion \cite{peres}. For the matrix $\rho$ of
Eq.~(\ref{eq:density_matrix}), (we set $\beta=\beta'$ for
simplicity) this involves looking if the matrix
\begin{equation}
  \label{eq:density_matrix pt}
  \rho_P=
  \left(
  \begin{array}{cccc}
  \alpha & 0 & 0 & 0 \\
 0 & \beta & \gamma & 0 \\
  0 & \gamma & \beta &0 \\
 0 & 0 & 0 & \alpha' \\
  \end{array}
  \right)\,.
\end{equation}
has negative eigenvalues.  If it does, the state is entangled;
otherwise \cite{h-iff} it is not. The state
\eqref{eq:density_matrix} is entangled  if and only if
$\gamma>\beta$. The reader will easily convince himself by
inspecting the tomographic data in \cite{nature} that this is
never the case for any of the measured states.


For the case at hand one can show that the measured states are all
unentangled even without the Peres criterion. For the sake of
simplicity we take $\alpha=\alpha'$ and $\beta=\beta'$. In this
case the matrix in Eq.~(\ref{eq:density_matrix}) can be written as
\cite{math}:
\begin{eqnarray}\label{brute}
    \rho&=& (\alpha-\gamma) \big(\rho_{\hat z}\otimes \rho_{\hat z}+
   \rho_{-\hat z}\otimes \rho_{-\hat z}\big)+
    (\beta-\gamma)\big(\rho_{\hat z}\otimes \rho_{-\hat z}+
   \rho_{-\hat z}\otimes \rho_{\hat z}\big) \nonumber\\
    &\phantom{=}&\,+\,\gamma \big(\rho_{\hat x}\otimes \rho_{\hat x}+
    \rho_{-\hat x}\otimes \rho_{-\hat x}
    +\rho_{-\hat y}\otimes \rho_{\hat y}+
    \rho_{\hat y}\otimes \rho_{-\hat y}\big)
\end{eqnarray}
where the  single qubit state $\rho_{\hat n}\ge 0$ is defined by
\begin{equation}\label{bla}
    \rho_n=
    \frac{\mathbb{I}+\hat n\cdot\vec\sigma}{2}\,
\end{equation}
$\vec\sigma=(\sigma_x,\sigma_y,\sigma_z)$ is the vector of Pauli
matrices; $\hat n$ a unit vector in three dimensions and
$\mathbb{I}$ the identity matrix.  Evidently, $\rho$ is a convex
combination of product states for all $\beta \ge \gamma$ and
$\alpha\ge \beta$ and agrees with the definition of a separable
(unentangled) state. Since this is the case for the tomographic
data for all the dots in \cite{nature} none of them produced an
entangled state.

As the above analysis shows, the emergence of off-diagonal terms
for dots with ``which path'' spectral ambiguity is not a satisfactory
evidence for entanglement as the authors of \cite{nature} claim.
The authors also subject their data to an additional tests of
entanglement. This test involves further processing of the data
that discards a significant part of the photon counts. Our
analysis  applies the definition of entanglement (alternatively
the Peres criterion) to the quantum state as measured by quantum
tomography. Thus any correct test for entanglement must agree with
our conclusions when applied to the same data.

In conclusion, either the quantum tomography data of the dots
studied in \cite{nature} is reliable and then none of the quantum
states produced in the experiment corresponds to entangled
photons; or the tomography data are not of sufficient quality and
no definite conclusion can be drawn from the experiment.

{\bf Note added:} After the above criticism  was communicated to
the authors of \cite{nature}, they have posted  a new paper on the
arXiv preprint server, describing results of a new experiment. In
this preprint they refer to \cite{nature} as the first
demonstration of emission of entangled photon pairs from quantum
dots. In light of the above, this claim is false. In fact, the
first correct demonstration can be found in \cite{ag}.\\

 {\bf Acknowledgment:} This work is supported
by the ISF.


\begin{thebibliography}{99}


\bibitem{nature} Stevenson, R. M., Young, R. J. Atkinson P., Cooper, K., Ritchi
e D. A., Shields A. J., A semiconductor source of triggered
entangled photon pairs, Nature {\bf 439} 179, (2006).

\bibitem{peres_book} Peres A., \textit{Quantum Theory: Concepts
and Methods}, Kluwer (Dordrecht) (1993).

\bibitem{kwiat}Kwiat, P.G., Waks, E., White, A.G., Appelbaum, L. Eberhard, P.H.,
Ultrabright source of polarization-entangled photons.
Phys.Rev.A{\bf60}, R773 (1999).

\bibitem{peres}Peres, A., Separability Criterion for Density Matrices, Phys.\ Rev.\ Lett. 77, 1413 (1996)
\bibitem{h-prl}M. Horodecki, P. Horodecki, and R. Horodecki,
Mixed-State Entanglement and Distillation: Is there a ¿Bound¿
Entanglement in Nature?,
 Phys. Rev. Lett. {\bf 80}, 5239¿5242 (1998)
\bibitem{h-iff}M. Horodecki, P. Horodecki, and R. Horodecki, Separability of mi
xed states:
necessary and sufficient conditions, Phys. Lett. A 223, 1 (1996),
quant-ph/9605038.
\bibitem{math} The following command lines compute the tensor
product of matrices in Mathematica:
\begin{verbatim}Needs["LinearAlgebra`MatrixManipulation`"]
KroneckerProduct[a_?SquareMatrixQ, b_?SquareMatrixQ]:=
BlockMatrix[Outer[Times,a,b]]
\end{verbatim}
\bibitem{ag} N. Akopian, N.H. Lindner, E. Poem. Y. Berlatzky,
J. Avron, D. Gershoni, D.B. Gerardot and P.M. Petroff, Entangled
photon pairs from a semiconductor quantum dot, quant-ph/0512048





\end{thebibliography}
\end{document}